\documentclass[conference,a4paper]{IEEEtran}
\usepackage[T1]{fontenc}
\pdfoutput=1

\makeatletter
\def\endthebibliography{%
	\def\@noitemerr{\@latex@warning{Empty `thebibliography' environment}}%
	\endlist
}
\makeatother

\IEEEoverridecommandlockouts
\usepackage{cite}
\usepackage{amsmath,amssymb,amsfonts}
\usepackage{algorithm}
\usepackage{algorithmic}
\usepackage{etoolbox}  
\usepackage{bbm}
\usepackage{amsmath}
\usepackage{amssymb}
\usepackage{amsthm}
\usepackage{bm}
\makeatletter
\patchcmd{\algorithmic}{\addtolength{\ALC@tlm}{\leftmargin} }{\addtolength{\ALC@tlm}{\leftmargin}}{}{}
\makeatother

\makeatletter
\newcommand\fs@betterruled{%
	\def\@fs@cfont{\bfseries}\let\@fs@capt\floatc@ruled
	\def\@fs@pre{\vspace*{5pt}\hrule height.8pt depth0pt \kern2pt}%
	\def\@fs@post{\kern2pt\hrule\relax}%
	\def\@fs@mid{\kern2pt\hrule\kern2pt}%
	\let\@fs@iftopcapt\iftrue}
\floatstyle{betterruled}
\restylefloat{algorithm}
\makeatother

\usepackage{filecontents}
\usepackage{graphicx}
\usepackage{textcomp}
\usepackage{xcolor}
\usepackage{authblk}
\def\BibTeX{{\rm B\kern-.05em{\sc i\kern-.025em b}\kern-.08em
		T\kern-.1667em\lower.7ex\hbox{E}\kern-.125emX}}

\begin{document}

\title{Latency and timeliness in multi-hop satellite networks}


\author{Beatriz~Soret$^{\dag}$, Sucheta Ravikanti$^{\S}$ and Petar Popovski$^{\dag}$\\
$^{\dag}$Department of Electronic Systems, Aalborg University, Denmark \\
$^{\S}$Department of Electrical Engineering, Indian Institute of Technology, Bombay}


\maketitle

\begin{abstract}
The classical definition of network delay has been recently augmented by the concept of information timeliness, or Age of Information (AoI). We analyze the network delay and the AoI in a multi-hop satellite network that relays status updates from satellite $1$, receiving uplink traffic from ground devices, to satellite $K$, using $K-2$ intermediate satellite nodes. The last node, $K$, is the closest satellite with connectivity to a ground station. The satellite formation is modeled as a queue network of M/M/1 systems connected in series. The scenario is then generalized for the case in which all satellites receive uplink traffic from ground, and work at the same time as relays of the packets from the previous nodes. The results show that the minimum average AoI is experienced at a decreasing system utilization when the number of nodes is increased. Furthermore, unloading the first nodes of the chain reduces the queueing time and therefore the average AoI. These findings provide insights for designing multi-hop satellite networks for latency-sensitive applications. \end{abstract}


\IEEEpeerreviewmaketitle

\section{Introduction}

\begin{figure}[t]
	\centering
	\includegraphics[width=3.8in]{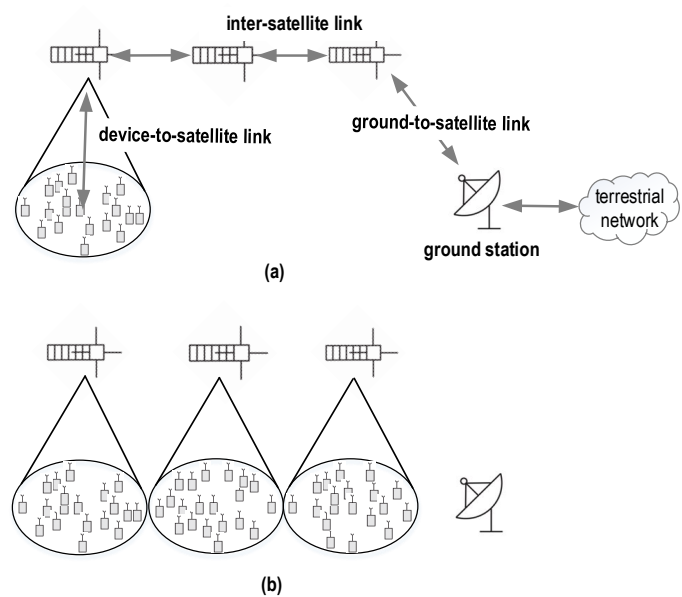}
	\caption{Example of multi-hop relaying satellite network. (a) Only the first satellite receives data from ground. (b) All satellites receive data from ground. }
	\label{fig_scenario}
\end{figure}

A well-known limitation of satellite communications is the inherent delay due to the large distances. Such propagation delay is highly reduced when using Low Earth Orbits (LEO), with altitudes between 500 and 2000 km, and propagation delays of few ms. Nevertheless, to ensure continuity in the coverage with a LEO satellite network, a flying formation of many satellites is required, usually organized in a \emph{constellation} with coordinated ground coverage. As a consequence, latency-sensitive information might suffer from long delays anyway, because several buffered-aided satellites are needed to connect two distant points on the Earth surface. The situation is illustrated in the example of Figure \ref{fig_scenario} (a). The first satellite provides coverage to a remote area. Particularly, we address the scenario in which status updates are received in the uplink from, e.g., a massive number of IoT devices, and this information must be relayed as soon as possible to the closest ground station.  Observe that the space segment works as a relay network where satellites are connected to each other via an inter-satellite link. The transmitted packet is downloaded by the closest satellite with a link to a ground station. A generalization is shown in Figure~\ref{fig_scenario}~(b), where all the satellites work as a relay and, at the same time, receive uplink status-updates from their coverage area. At each satellite $k$, the inter-satellite link is used to forward both the ground information from satellite $k$ and the packets from satellites $1..k-1$. 

\begin{figure*}[t]
	\centering
	\includegraphics[width=6in]{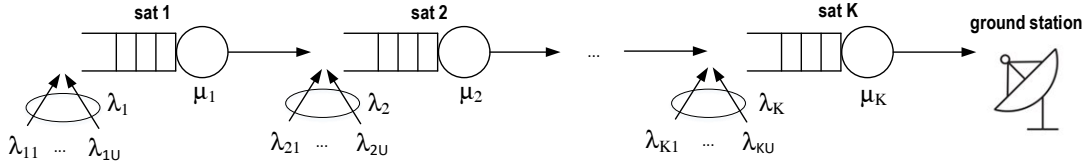}
	\caption{System model.}
	\label{fig_system_model}
\end{figure*}

In multi-hop relay networks, the introduction of intermediate nodes from transmitter to receiver has the drawback of additional latency \cite{Khanna2018} \cite{Yang2018}. Beyond traditional network delay measures, the concept of Age of Information (AoI) was introduced in \cite{Kaul2012} to better capture the timeliness of the received information. AoI measures the time that elapsed since the last received fresh update was generated at the source. Pertinent applications are those where a source generates updates that are transmitted through a communication network \cite{Kaul2011}, like common satellite services that involve a certain tracking, such as containers in logistics. This is also relevant for Authentication Identification System (AIS) data from ships/vessels or the Automatic Dependent Surveillance - Broadcast (ADS-B) system in airplanes, as the end receiver is interested in the freshest update. 

The AoI metric has been studied under various conditions. The original paper from Kaul \cite{Kaul2012} analyzed a single G/G/1 queue with exponential and deterministic distributions. The optimization based on AoI metrics of both the network and the senders’ updating policies has been discussed in \cite{Sun2017} and \cite{Yates2015}. In \cite{Talak2017}, the problem of multi-hop networks with many source-destination pairs and interference constraints is addressed, and the optimal policy is reduced to solving the equivalent problem in which all source-destination pairs are just a single-hop away. \cite{Kuang2019} studies the average AoI of a tandem queue system (i.e., only two nodes), with a zero-waiting policy in the second node. This does not fit the satellite scenario in which traffic enters also from the intermediate nodes. In this paper, we address buffer-aided multi-hop networks like the ones depicted in Figure~\ref{fig_scenario}, used to relay information between two remote ground points. Imagine we have two satellites and only the second one has connectivity to a ground station. In that case, if all the uplink ground traffic is received by the first satellite, then all packets are queued up twice, in the first and in the second satellite, respectively. In the other extreme, when the ground traffic is only received by the second satellite, then the packets are queued up only in the second buffer. Our analysis captures the intermediate cases as well, and provides insights for designing constellations for latency-sensitive satellite services.   

In the rest of the paper, Section \ref{sec:systemmodel} presents the system model, Section \ref{sec:AoInointertraffic} analyzes the average AoI when only the first node receives uplink traffic and the rest of nodes are just relays, Section \ref{sec:AoIintertraffic} generalizes to the case in which all satellites cover populated areas, and Section \ref{sec:conclusions} discusses the conclusions and future directions of this work.

\section{Setting the scene} \label{sec:systemmodel}

The satellite network consists of a total of $K$ satellites (3 in Figure \ref{fig_scenario}) connected with an inter-satellite link (ISL). We assume the uplink traffic from remote areas is transmitted to the space segment using the first available satellite. If the information is delay-sensitive (e.g., status updates), then it must be routed as soon as possible to the destination, which is the closest satellite with current connection to a ground station or a ground gateway. The number of ground stations in satellite networks is often limited to one or few of them across the globe, and therefore it is common that several hops are required to reach the destination. The system is hence a multi-hop relay network where the last satellite in the group has a reachable ground station, and the other ones use the ISL to route the data. Each relay gets packets from the previous node and from ground, except for the first node. 

This multi-hop satellite network is modeled as a queueing network connected in series, as shown in Figure \ref{fig_system_model}. 
Satellite $k$ collects and buffers packets from a total of $U_k$ ground devices located in its coverage area. If the load is equally shared among satellites, then $U_k = U \; \forall k$. The first node has only ground traffic, but the intermediate nodes receive both traffic from the previous node, to be relayed to the next one, and the own traffic from ground. The last node, $K$, has connection to a ground station, and  downloads all the received traffic to the Earth. The transmission is ideal, with no packet drops, i.e., a node $k$ receives all reports from node $k-1$, which are forwarded to node $k+1$. Notice that traffic from ground and from the ISL are stored in the same buffer, with no priorities among them. Each node is equipped with a buffer of infinite capacity, ruled by a First Come First Served (FCFS) policy with no priorities among packets or links.

The servers model the ISL between nodes, i.e., the wireless transmission, for nodes $k=1..K-1$. In the last node, $K$, the server models instead the downlink to ground. In this initial study all the links are assumed to be equal, with a service rate $\mu=\mu_k,\; k=1, ..., K$.  The same service time means that all the links are equally characterized. The obvious extension is to consider the case in which the service rate $\mu_K$ of the satellite-to-ground link is different from the service rates of the inter-satellite links; this is left for future work\footnote{The difference is that in our analysis we sum exponential random variables of the same rate, which gives an Erlang distribution. If each link has its own rate, then the distribution is a Hypoexponential}.  The service time of each satellite is the inverse of the service rate, $S_k = 1/\mu_k$. 

From ground, each device generates status updates at a Poisson rate $\lambda_{ku}$. The aggregated traffic received at node $k$ is the sum of the traffic received from $U_k$ devices and the traffic received via the ISL with the previous node, and it follows a Poisson distribution as well. The rate at each node is
\begin{equation}
\lambda_{k} = \left\{
\begin{array}{l}
\sum_{u=1}^{U_k}\lambda_{ku}, \quad k = 1\\
\sum_{u=1}^{U_k}\lambda_{ku} + \lambda_{k-1}, \quad k > 1\\
\end{array}
\right.
\end{equation}

\section{Average Age of Information with ground traffic in the first node} \label{sec:AoInointertraffic}

\subsection{Analysis with $K$ nodes}
We consider a satellite multi-hop network of $K$ nodes like in Figure \ref{fig_scenario} (a). The first satellite covering a remote area receives Poisson traffic at an aggregated rate $\lambda_1$ from ground, and this traffic is relayed using nodes $2, 3, ..., K$. In this case, the arrival rate simplifies to \mbox{$\lambda = \lambda_{1} = \sum_{u=1}^{U} \lambda_{1u}$} and \mbox{$\lambda_{k} = \lambda_{k-1}, \;\;\; k > 1$} (no lost packets). 

An update is said to be fresh when its timestamp is the current time $t$ and its age is zero. The age at the destination (the ground station) increases linearly in time in the absence of any updates, and is reset to a smaller value when an update is received. Define the AoI in the destination node $K$ at time $t$ as the random process $\Delta(t) = t - u(t)$.

\begin{figure}[t]
	\centering
	\includegraphics[width=3.6in]{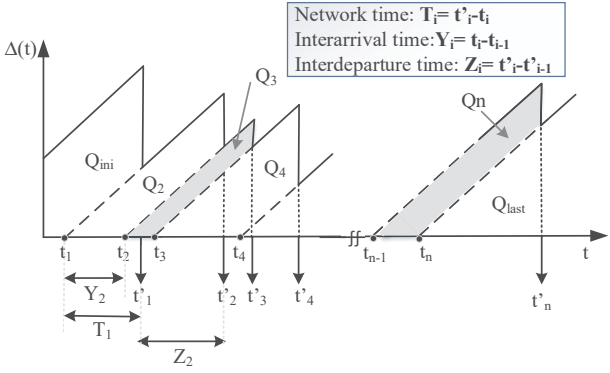}
	\caption{Evolution of the Age of information in a queue network with $K$ nodes. The network times $T_i$ are defined as the total time spent in the system, since arrival in node $1$ until departure in node $K$.}
	\label{fig:aoi_FCFS}
\end{figure}

The evolution of the AoI $\Delta(t)$ at the destination under a FCFS policy exhibits the sawtooth pattern plotted in Figure \ref{fig:aoi_FCFS}. Without loss of generality, the system is first observed at $t=0$ and the queue is empty with $\Delta(0) = \Delta_0$. Index $i$ is for packet $i$. Status update $i$ is generated at time $t_i$ and is received by the destination at time $t_i'$. Define $Y_i$ as the interarrival time $Y_i = t_i - t_{i-1}$ between two packets; $Z_i$ as the interdeparture time $Z_i = t'_i - t'_{i-1}$; and $T_i$ as the total network time in the system $T_i = t'_i - t_i$. The latter includes the time spent in all the nodes (queueing time and transmission time) until departure from the system at node $K$. This is different from \cite{Kuang2019}, where the times are defined within a single buffered system. In our case, the network is as a connection of M/M/1 systems, where only the first one gets external load. 

To evaluate the average AoI, the strategy is to calculate the area under $\Delta(t)$, or the time average AoI, as
\begin{equation}
    \Delta_{\mathcal{T}} = \frac{1}{\mathcal{T}} \left(Q_{\text{ini}} + Q_{\text{last}}+\sum_{i=2}^{N(\mathcal{T})} Q_i \right)
\end{equation}

\noindent where $N(\mathcal{T})$ is the number of arrivals by time $\mathcal{T}$. The average AoI $\bar{\Delta}$ is given by the limit
$\bar{\Delta} = \lim_{\mathcal{T} \rightarrow \infty} \Delta_{\mathcal{T}} $.

As defined in Figure \ref{fig:aoi_FCFS}, $Q_i(i>1)$ are trapezoids whose areas can be calculated as the differences between two isosceles triangles \cite{Kuang2019}, i.e., 
\begin{align}
Q_i = \frac{1}{2}\left(T_i + Y_i\right)^2 - \frac{1}{2}T_i^2 = Y_i T_i + \frac{Y_i^2}{2}
\end{align}

The average AoI $\bar{\Delta}$ can be expressed
\begin{align}
\bar{\Delta} &= \lambda E[Q_i] =
\lambda \left( E[T_i Y_{i}] +
E\left[\frac{1}{2}Y_{i}^2\right]\right) \label{eq:aoi}
\end{align}

\noindent where $\lambda = \lim_{\mathcal{T} \rightarrow \infty}\frac{N(\mathcal{T})}{\mathcal{T}}$ is the steady state rate of status updates generation. Ergodicity has been assumed for the stochastic process $\Delta(t)$, but not assumptions regarding the distribution of the random variables $Y$ and $T$ have been made. 
	
The second term in equation (\ref{eq:aoi}), $E\left[\frac{1}{2}Y_{i}^2\right]$, is easily obtained as
\begin{equation}
E\left[\frac{1}{2}Y_{i}^2\right] = \frac{1}{\lambda^2} \label{eq:second_term}
\end{equation}

For the first term, $E[T_i Y_{i}]$, notice that the total system time of packet $i$ is the sum of the system times in each of the nodes $1, 2, ...K$, and each of them can be decomposed in waiting and service time
\begin{equation}
T_i = W_{i,1} + S_{i,1} + W_{i,2} + S_{i,2} + ... + W_{i,K} + S_{i,K}
\end{equation}

We rewrite this term to get
\begin{align}
E[T_iY_{i}] &= E[(W_i + S_i) Y_{i}] = E[W_i Y_{i}] + E[ S_i]E[Y_{i}] \nonumber \\ &= E[W_i Y_{i}] + \sum_{k=1}^{K}E[ S_{i,k}]E[Y_{i}] \label{eq:tiyi}
\end{align}

The whole waiting time in the system is given by $W_{i,1}+...+W_{i,k}$. When packet $i$ arrives to the system there are two scenarios. Let us take the simplest case of two nodes. The first situation is that packet $i-1$ has already left the system when $i$ is generated, then $W_{i} = 0$. In the second possibility, packet $i-1$ leaves the system after the arrival of packet $i$. It can happen that when packet $i-1$ leaves the system, packet $i$ is in the first server; or it is waiting in the second queue. If packet $i$ does not find the last queue empty, we can write the total waiting time of packet $i$ in the general case of $K$ nodes as:  
\begin{equation}
W_{i} = \left(T_{i-1} - Y_{i} - \sum_{k=0}^{K-1}S_{i,k}\right)^+ = \left(T_{i-1} - Y_{i} - S_{\setminus K}\right)^+ 
\label{eq:wi}
\end{equation}

\noindent where we have defined $S_{\setminus K}=\sum_{k=1}^{K-1}S_{i,k}$. Notice also that the time in equation (\ref{eq:wi}) is the total time spent in queues across the network, one or all of them, assuming a non-empty last queue. 

When the system reaches steady state the system times are stochastically identical, that is $T = ^{st} T_i =^{st} T_{i-1}$. 

We can now write the conditional expected waiting time,  

\begin{align}
E[W_i | Y_i = y+S_{\setminus K}] &= E[(T-y-S_{\setminus K})^+] \label{eq:conditional_nointer}
\end{align}

To solve equation (\ref{eq:conditional_nointer}), we need the distribution of the system time. If the system is not saturated, i.e., if the server utilization in each M/M/1 meets the condition $\rho_k = \lambda/\mu_k < 1$, then Burke's theorem can be applied \cite{Burke1956}. This means that the departure process from satellite $k$ is also Poisson.  


The service rates are equal across the network and, under no saturation, the arrival rate at queue $k$ is the same as queue $k-1$. Therefore, we can conclude that the server utilization is also equal for all nodes and define $\rho_k = \rho = \frac{\lambda}{\mu} \;\; \forall k$. 
We define $\alpha = \mu (1-\rho)$ to be the parameter of the exponential distribution of a single M/M/1 stage, $f_{T_{M/M/1}}(t) = \alpha t e^{-\alpha t}$.

With $K$ nodes, the probability distribution function (pdf) of the total network time is the sum of i.i.d. exponential variables of the same rate $\alpha$, i.e., an Erlang distribution
\begin{equation}
f_T(t) = \frac{\alpha^K t^{K-1} e^{-\alpha t}}{(K-1)!} \label{eq:sojourn_distribution}
\end{equation}

The mean of the distribution is the network delay, $E[T]$, given by \cite{Burke1956}
\begin{equation}
E[T] = \frac{K}{\alpha}  \label{eq:network_delay}
\end{equation}

The conditional expected waiting time is
\begin{align}
&E[W_i | Y_i = y, S_{\setminus K}=s] = E[(T-y-s)^+] \nonumber \\
=& \int_{y+s}^{\infty} (t-y-s)\frac{\alpha^K t^{K-1}}{(K-1)!} e^{-\alpha t} dt  \nonumber\\
&= \frac{\Gamma(K+1, \alpha(y+s))}{\alpha \Gamma(K)} - \frac{(y+s) \cdot \Gamma(K, \alpha(y+s))}{\Gamma(K)}
\label{eq:conditional}
\end{align}  

\noindent where $\Gamma(s,x) = \int_x^{\infty}{t^{s-1}e^{-t} dt}$ is the upper incomplete gamma function, and $\Gamma(s) = \int_0^{\infty}{t^{s-1}e^{-t} dt}$ is the ordinary gamma function.

We apply the law of total probability to uncondition equation (\ref{eq:conditional}) on $Y_i$, and use the exponential interarrival time \mbox{$f_{Y_i}(y) = \lambda e^{-\lambda y}$} to obtain the expectation 
\begin{equation}
  E[W_i Y_i | S_{\setminus K}=s]= \int_0^{\infty} y \mathbb{E}\left[W_i | Y_i = y, S_{\setminus K}=s\right] f_{Y_i}(y) dy . \label{eq:WIgivenS}  
    \end{equation}
    
We notice that by using the average $E[S_{\setminus K}] = \frac{K-1}{\mu}$ instead of unconditioning equation (\ref{eq:WIgivenS}) on $S_{\setminus K}$, we obtain a lower bound of the desired expectation $E[W_i Y_i]$, with the result given in equation (\ref{eq:I1}). 

\begin{figure*}[t]
\begin{align}
E[W_i Y_i]  
&\leq \frac{-1}{\alpha \lambda^2 \Gamma(K)} \left( \alpha(\lambda S_{\setminus K}+2) \Gamma(K, \alpha S_{\setminus K}) - \lambda \Gamma(K+1, \alpha S_{\setminus K}) \right)  \nonumber
\\ 
&- \frac{ \alpha^K \mu^{-K} e^{\lambda S_{\setminus K}}}{\lambda^2 \mu \Gamma(K)}  
\cdot \left( \mu(\lambda S_{\setminus K}-2) \Gamma(K, \mu S_{\setminus K} ) - \lambda \Gamma(K+1, \mu S_{\setminus K}) \right)  \label{eq:I1}
\end{align}
\end{figure*} 

The last term in equation (\ref{eq:tiyi}) is easily obtained as $\sum_{k=1}^{K}E[ S_{i,k}]E[Y_{i}] = \frac{K}{\mu \lambda}$.

And the expectation $E[T_iY_{i}]$ yields
\begin{equation}
E[T_iY_{i}] = E[W_i Y_i]  + \frac{K}{\mu \lambda}
\end{equation}

The average AoI for $K$ nodes can be finally bounded by
\begin{align}
\bar{\Delta} &= 
\lambda \left( E[W_i Y_i]  +  \frac{K}{\mu \lambda} + \frac{1}{\lambda^2}
\right) \label{eq:aoi2_final}
\end{align}
\noindent where $E[W_i Y_i]$ is lower bounded by equation (\ref{eq:I1}). 

\subsection{Numerical evaluation}
\begin{figure}[t]
	\centering
	\includegraphics[width=3.8in]{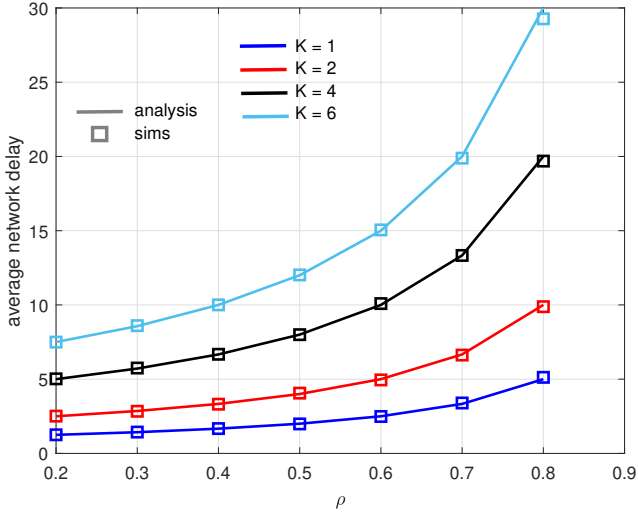}
	\caption{Network delay for $K$ nodes versus the system utilization. $\mu_k = 1\;\forall k$. } 
	\label{fig:system_delay}
\end{figure}

The results are numerically evaluated and compared to MATLAB simulations of the system. First, Figure \ref{fig:system_delay} shows the network delay of equation (\ref{eq:network_delay}) as a function of the system utilization load $\rho$ and for various values of $K$. The baseline case $K=1$ is also plotted for reference. As expected, $E[T]$ increases with the system utilization and the number of nodes.

\begin{figure}[t]
	\centering
	\includegraphics[width=3.8in]{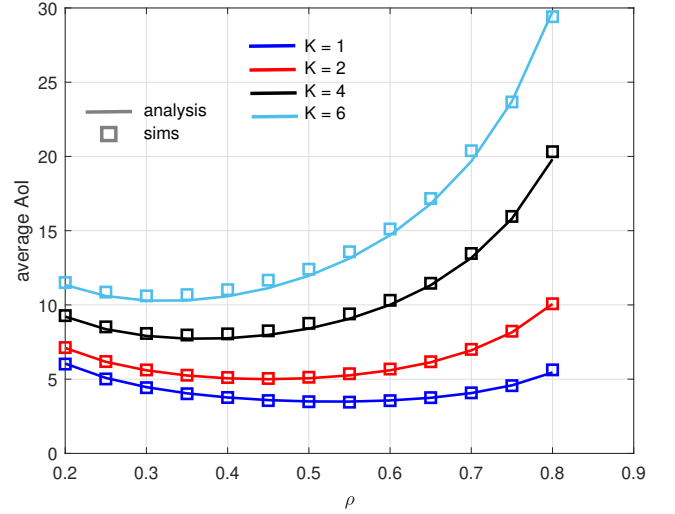}
	\caption{Average AoI for $K$ nodes versus the system utilization. $\mu_k = 1\;\forall k$. }
	\label{fig:results_no_errors}
\end{figure}

More interestingly, Figure \ref{fig:results_no_errors} plots the lower bound of the average AoI for the same setting. $\bar{\Delta}$ increases with the number of relaying nodes. The baseline case $K=1$ was first provided in \cite{Kaul2012}. The lower bound is very tight and it approximates almost perfectly the simulation results. This indicates that the variability of the service time has not a big impact in $\bar{\Delta}$. For fixed service rate $\mu$, we can minimize $\bar{\Delta}$ with respect to the arrival rate $\lambda$ or, equivalently, to the system utilization $\rho$. In the case of $K=1$ this minimum is attained at $\rho^* \approx 0.53$ \cite{Kaul2012}. As $K$ increase, the minimum $\rho^*$ diminishes, as shown in Figure \ref{fig:rho}. For example, with $10$ hops the optimal age is achieved by choosing a $\lambda$ that biases the server towards being busy less than $30$ \% of the time.

\section{Average Age of Information with ground traffic in all nodes} \label{sec:AoIintertraffic}

We extend the scenario in Section \ref{sec:AoInointertraffic} to the case in which the intermediate nodes do receive traffic from ground (Figure \ref{fig_scenario} (b)), i.e., \mbox{$\lambda_{k} = \sum_{u=1}^{U}\lambda_{ku} + \lambda_{k-1}, \;\;\; k > 1$}. The analysis considers the total network traffic, therefore it is possible that packet $i$ is received by satellite $k$ before packet $i-1$ because packet $i$ is received by a satellite which is closer to $k$.

\subsection{Analysis with two nodes}

Let us examine the simplest case of two nodes, $K = 2$. We define $\lambda = \lambda_1 + \lambda_2$ to be the total arrival rate to the system, and $p$ to be the proportion of arrival rate in the first node, i.e., $\lambda_1 = p \lambda$ and $\lambda_2 = (1-p) \lambda$. Therefore, the utilization in the second satellite is limited to $\rho_2 = (\lambda_1 + \lambda_2)/\mu = \lambda/\mu = \rho$. 

To address this case, we notice that a packet leaving the system has a probability $p$ of coming originally from the first node, and a probability $1-p$ of coming from the second node. Therefore, we calculate the average age as
\begin{equation}
    \bar{\Delta} = p \bar{\Delta}_{\text{node 1}} + (1-p) \bar{\Delta}_{\text{node 2}} \label{eq:AoI_p}
\end{equation} 
\noindent where we define $\bar{\Delta}_{\text{node 1}}$ as the average AoI of a packet that arrives at node 1, and $\bar{\Delta}_{\text{node 2}}$ for a packet arriving at node 2. 

The packet arriving at node 2 sees the system as a single M/M/1, and therefore $\bar{\Delta}_{\text{node 2}}$ is written as the average AoI of an M/M/1 system with a total arrival rate $\lambda$, given in \cite{Kaul2012} as
\begin{equation}
    \bar{\Delta}_{\text{node 2}} = \lambda \left( \frac{\rho}{\mu^2(1-\rho)} + \frac{1}{\mu \lambda} + \frac{1}{\lambda^2} \right)
\end{equation}

Instead, a packet arriving to the first node finds two M/M/1 systems connected in series. The two M/M/1 stages are not equal, as it was the case in Section \ref{sec:AoInointertraffic}, because the load in the second node is higher. This changes the distribution of the network time and therefore the AoI. The pdf of the network time is obtained as the pdf of the sum of two independent exponential variables with different rate $\alpha_k$, i.e.,
\begin{equation}
   f_T(t) = \frac{\alpha_1 \alpha_2}{\alpha_1 - \alpha_2} (e^{-\alpha_2 t} - e^{-\alpha_1 t}) \label{eq:sojourn_2}
\end{equation}

\noindent where $\alpha_k = \mu_k(1-\rho_k)$.

Like in Section \ref{sec:AoInointertraffic}, we calculate the conditional expected waiting time in this new scenario, which yields
\begin{equation}
E[W_i | Y_i = y, S_{\setminus 2}=s]
= \frac{\alpha_1 e^{-\alpha_2(y+s)}}{\alpha_2(\alpha_1 - \alpha_2)}
- \frac{\alpha_2 e^{-\alpha_1(y+s)}}{\alpha_1(\alpha_1 - \alpha_2)}
\label{eq:conditional22}
\end{equation}

In this two-nodes case, $S_{\setminus 2} = S_1$, i.e., the sum of all the service times except the last node is just the service time of the first node. The expectation $E[W_i Y_i]$ for these packets arriving to the first satellite is lower bounded by
\begin{equation}
E[W_i Y_i ] \leq - \frac{\alpha_1 \lambda e^{-\alpha_2 S_{\setminus 2}}}{\alpha_2(\alpha_1 - \alpha_2)(\alpha_2 + \lambda)^2}
+ \frac{\alpha_2 \lambda e^{-\alpha_1 S_{\setminus 2}}}{\alpha_1(\alpha_1 - \alpha_2)(\alpha_1 + \lambda)^2}
\label{eq:conditional2}
\end{equation}

The average AoI of packets arriving at node 1 is
\begin{equation}
    \bar{\Delta}_{\text{node 1}} = \lambda \left( E[W_i Y_i] + \frac{2}{\mu \lambda} + \frac{1}{\lambda^2}  \right)
\end{equation} 
\noindent where $E[W_i Y_i]$ is given by equation (\ref{eq:conditional2}). 

The mean network delay, $E[T]$, is calculated as
\begin{equation}
E[T] = p \left(\frac{1}{\alpha_1} +  \frac{1}{\alpha_2}\right) + (1-p) \left(\frac{1}{\alpha_2} \right) \label{eq:sojourn_distribution2}
\end{equation}

The two extreme cases of this two-node scenario are easily obtained. When $p = 1$, all the ground load comes from the first satellite, and the AoI is given by equation (\ref{eq:aoi2_final}) with $K=2$. When $p=0$, all the load is in the second satellite, and the system reduces to an M/M/1 whose AoI is given in \cite{Kaul2012}.

\begin{figure}[t]
	\centering
	\includegraphics[width=3.8in]{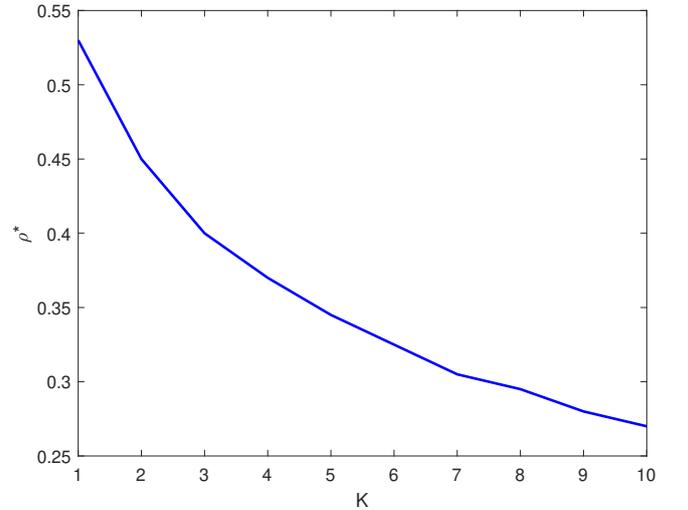}
	\caption{System utilization that minimizes the AoI for increasing number of nodes.}
	\label{fig:rho}
\end{figure}
\subsection{Numerical evaluation}
\begin{figure}[t]
	\centering
	\includegraphics[width=3.8in]{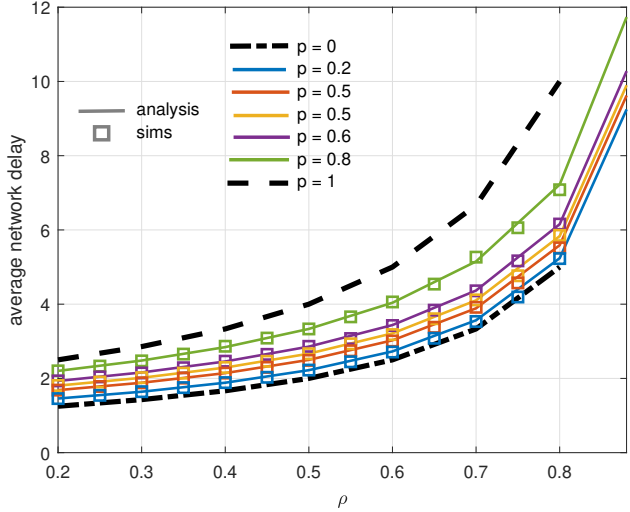}
	\caption{Network delay for $2$ nodes each of them receiving traffic from ground, versus the system utilization.  $\mu_k = 1\;\forall k$; $\lambda_1 = p \lambda$; $\lambda_2 = (1-p) \lambda$. } 
	\label{fig:system_delay_inter}
\end{figure}

\begin{figure}[t]
	\centering
	\includegraphics[width=3.8in]{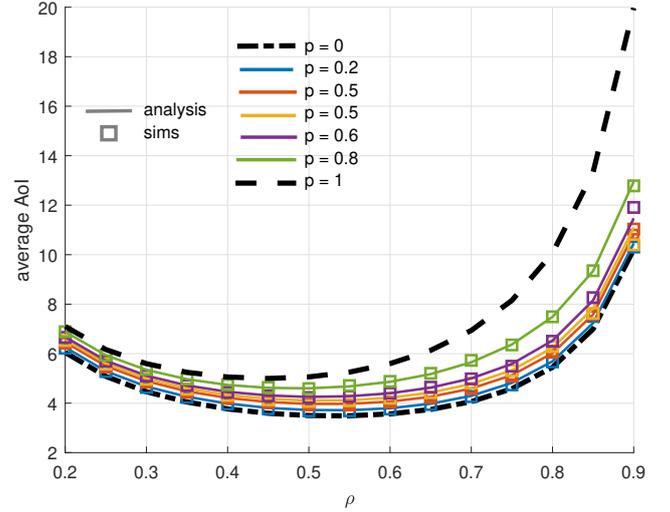}
	\caption{Average AoI for $2$ nodes each of them receiving traffic from ground, versus the system utilization. $\mu_k = 1\;\forall k$; $\lambda_1 = p \lambda$; $\lambda_2 = (1-p) \lambda$}
	\label{fig:inter_traffic_p}
\end{figure}

Figures~\ref{fig:system_delay_inter} and \ref{fig:inter_traffic_p} show the network delay in equation (\ref{eq:sojourn_distribution2}) and the average AoI of equation (\ref{eq:AoI_p}) for two nodes, increasing values of $\rho$ and different values of $p$. The simulated results and the two bounds $p = 0$ and $p = 1$ are also plotted for reference. As expected, the average AoI $\bar{\Delta}$ in Figure \ref{fig:inter_traffic_p} is within the two bounds. When the system utilization is low, many packets do not experience queueing in the first and second satellite (i.e., the server is empty), and therefore the differences in delays and AoI is reduced to the service times. Since we are using a normalized value $\mu = \mu_k = 1$, the minimum difference between queueing up in one or two buffers is 1. This is visible, e.g., in Figure \ref{fig:inter_traffic_p}, where the difference between the curves of $p = 0$ and $p = 1$ converges to $1$ as $\rho$ tends to zero. As $\rho$ increases, the positive impact of moving devices to the second node is more noticeable. For instance, moving 20\% of the total load to the second node reduces the average AoI by $\approx$10\% at $\rho = 0.5$ (i.e., the server is idle half of the time), and by $\approx$35\% at $\rho = 0.9$. Also, $\rho^*$ decreases slowly as $p$ is increased. 

\begin{figure}[t]
	\centering
	\includegraphics[width=3.8in]{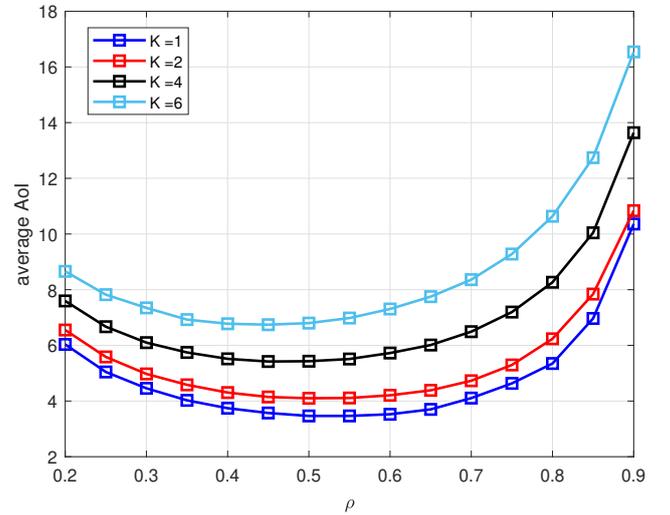}
	\caption{Simulated average AoI for $K$ nodes each of them receiving traffic from ground, versus the system utilization. $\mu_k = 1\;\forall k$}
	\label{fig:inter_traffic_K}
\end{figure}

Finally, Figure~\ref{fig:inter_traffic_K} plots the simulation evaluation of $\bar{\Delta}$ for increasing values of $\rho$ and $K$, when all the nodes are equally loaded from ground, i.e., $\lambda_1 = \lambda_2 = ...$. Naturally, the last node is the bottleneck, because it receives the traffic from ground and the aggregated traffic to be relayed from satellites $1, 2, ... K-2$. $\Delta$ increases as the number of satellite nodes gets larger, and the system utilization that minimizes the AoI is also decreased with $K$, similarly as the behaviour without traffic in the intermediate nodes that was shown in Figure \ref{fig:rho}. 

\section{Conclusions} \label{sec:conclusions}
In this paper, we have analyzed the average AoI in a multi-hop satellite network, where status updates have to be relayed from satellite $1$ to satellite $K$ with connectivity to a ground station. We have generalized the AoI definition to include a connection of M/M/1 queues, and quantified the increase in average AoI when the number of satellite nodes grows. We then extend the study to address the situation in which each of the satellite in the network can receive status updates from ground, and therefore the last satellite in the sequence is the most congested node. The performance evaluations in terms of number of nodes, distribution of the load among satellites and freshness reveals interesting tradeoffs that are relevant to the design and dimensioning of multi-hop satellite networks. The system utilization (or, equivalently, the arrival rate) that minimizes the average AoI decreases as the number of hops increases. The influence of the packet drops and the tradeoffs between reliability and freshness are left for future work. 

\section*{Acknowledgment}

This work has been in part supported by the European Research Council (ERC Consolidator Grant no. 648382 WILLOW). The authors would like to thank Kristian J. Tilsted and Federico Chiariotti from Aalborg University for their help. 

\bibliographystyle{IEEEtran}
\bibliography{ICC_AoI_relay}

\end{document}